\documentclass[alpha-refs]{wiley-article}

\usepackage{setspace}
\usepackage{lineno}

\usepackage[english]{babel}
\usepackage[T1]{fontenc}

\usepackage{amsmath}
\usepackage{graphicx}
\usepackage[colorinlistoftodos]{todonotes}
\usepackage[colorlinks=true, allcolors=blue]{hyperref}

\usepackage{natbib}

\usepackage{rotating}

\title{Automatic acoustic detection of birds through deep learning: the first Bird Audio Detection challenge}

\author[1]{Dan Stowell}
\author[2]{Mike Wood}
\author[3]{Hanna Pamu{\l}a}
\author[4]{Yannis Stylianou}
\author[5]{Herv\'e Glotin}

\affil[1]{Machine Listening Lab, Centre for Digital Music, Queen Mary University of London.}
\affil[2]{Ecosystems and Environment Research Centre, School of Environment and Life Sciences, University of Salford.}
\affil[3]{Department of Mechanics and Vibroacoustics, AGH University of Science and Technology, 30-059 Kraków}
\affil[4]{Computer Science Department, University of Crete.}
\affil[5]{LSIS UMR CNRS, University of Toulon, Inst. Univ. de France}
\corraddress{Dan Stowell, Machine Listening Lab, Centre for Digital Music, Queen Mary University of London, London, UK.}
\corremail{dan.stowell@qmul.ac.uk}
\fundinginfo{
EPSRC fellowship EP/L020505/1;
NERC grant NE/L000520/1;
ERASMUS+;
AGH Research Grant No 15.11.130.642
}
\runningauthor{Stowell et al., Automatic acoustic detection of birds}

\begin{document}
\maketitle

\begin{abstract}%
Assessing the presence and abundance of birds is important for monitoring
specific species as well as overall ecosystem health. Many birds are
most readily detected by their sounds, and thus passive acoustic monitoring 
is highly appropriate.
Yet acoustic monitoring is often held back by practical limitations
such as the need for manual configuration, reliance on example
sound libraries, low accuracy, low robustness, and limited ability to
generalise to novel acoustic conditions.
Here we report outcomes from a collaborative data challenge showing
that with modern machine learning including deep learning, general-purpose acoustic bird
detection can achieve very high retrieval rates in remote monitoring
data --- with no manual recalibration, and no pre-training of the
detector for the target species or the acoustic conditions
in the target environment. Multiple methods were able to attain
performance of around 88\% AUC (area under the ROC curve), much higher
performance than previous general-purpose methods.
We present new acoustic monitoring datasets,
summarise the machine learning techniques proposed by challenge teams,
conduct detailed performance evaluation,
and discuss how such approaches to detection can be integrated into remote monitoring
projects.
\end{abstract}


%%%%%%%%%%%%%%%%%%%%%%%%%%%%%%%%%%%%%%%%%%%
\newpage
\section{Introduction}

Worldwide, bird populations have exhibited steep declines since the 1970s, largely due to changes in land management \cite{RSPB:2013,Sonab:2016}.
Bird populations are also expected to change in number and distribution as the impacts of climate change play out in coming years \citep{Johnston:2013}.
It is thus crucial to monitor avian populations, for the purposes of conservation, scientific research and ecosystem management.
This has traditionally been performed via manual surveying,
often including the use of volunteers to help address the challenges of scale \citep{Johnston:2014,Kamp:2016}.
However, manual observation remains limited, especially in areas that are physically challenging to access, or night-time behaviour.
Many bird species are readily detectable by their sounds, often more so than by vision,
and so with modern remote monitoring stations able to capture continuous audio recordings
the prospect opens up of massive-scale spatio-temporal monitoring of birds
\citep{Aide:2013,Furnas:2015,Hill:2017,Matsubayashi:2017,Frommolt:2017,Knight:2017}.

The first wave of such technology performs automatic recording but not automatic detection,
relying on manual after-the-fact study of sound recordings \citep{Furnas:2015,Frommolt:2017}.
Later projects have employed some form of automatic detection,
which might be based on low-complexity signal processing such as energy thresholds
or template matching \citep{Towsey:2012,Colonna:2015}, or on machine learning algorithms \citep{Aide:2013}.
However, when used for field deployments, practitioners face a common hurdle.
With the current state of the art, all methods require manual tuning of algorithm parameters, customisation of template libraries and/or post-processing of results,
often necessitating some degree of expertise in the underlying method.
The methods are not inherently able to generalise to new conditions---whether those conditions be differing species balances, noise conditions, or recording equipment.
Many methods also exhibit only moderate accuracy,
which is tolerable in small surveys but leads to unfeasible amounts of false-negatives and -positives in large surveys \citep{Marques:2012}.
A further common limitation is the lack of robustness in particular to weather noise:
sounds due to rain and wind are commonly observed to dramatically affect detector performance,
and as a result surveys may need to treat weather-affected recording periods as missing data.

Recent decades have witnessed extremely strong growth in the abilities of machine learning.
The advances are due to increased dataset sizes and computational power but also due to \textit{deep learning} methods that can learn to make predictions in extremely nonlinear problem settings, such as speech recognition or visual object recognition \citep{Lecun:2015}.
These methods have indeed been applied to bioacoustic audio tasks \citep{Goeau:2016,Salamon:2017,Knight:2017},
and it is clear that their use could enable many organisations to work more cheaply and efficiently \citep{Joppa:2017}.
However, even with the strong performance of modern machine learning there remain important questions about generalisability \citep{Knight:2017}.
Machine learning workflows use a ``training set'' of data from which the algorithm ``learns'', optionally a ``validation set'' used to determine when the learning has achieved a satisfactory level, and then a ``testing set'' which is used for the actual evaluation, to estimate the algorithm's typical performance on unseen data.
Such evaluation is typically performed in \textit{matched conditions}, meaning the training and testing sets are drawn from the same pool of data,
and thus general properties of the datasets---such as the number of positive versus negative cases---are expected to be similar.
This enables users to test that the algorithm can generalise to new items drawn from the same distribution.
However, in practical deployments of machine learning the new items are rarely drawn from the same distribution:
conditions drift, or the tool is applied to new data for which no training data are available \citep{Sugiyama:2012,Knight:2017}.
This is one reason that accuracy results obtained in research papers might not translate to the field.

In order to address such problems, we designed a public evaluation campaign focused on a highly-general version of the bird detection task,
intended specifically to encourage detection methods which are able to generalise well:
agnostic to species, and able to work in unseen acoustic environments.
In this work we present the new acoustic datasets we collated and annotated for the purpose,
the design of the challenge, and its outcomes, with new deep learning methods able to achieve strong results despite the difficult task.
We analyse the submitted system outputs for their detection ability as well as their robust calibration,
we perform a detailed error analysis to inspect the sound types that remain difficult for machine learning detectors,
and apply the leading system to a separate held-out dataset of night flight calls.
We conclude by discussing the new state of the art represented by the deep learning methods that excelled in our challenge,
the quality of their outputs and the feasibility of deployment in remote monitoring projects.

%%%%%%%%%%%%%%%%%%%%%%%%%%%%%%%%%%%%%%%%%%%
\section{Materials and methods}

To conduct the evaluation campaign we designed a detection task to be solved---specific but illustrative of general-purpose detection issues---%
gathered multiple datasets and annotated them, and then led a public campaign evaluating the results submitted by various teams.
After the campaign, we performed detailed analysis of the system outputs, inspecting questions of accuracy, generality and calibration.

Our aim to facilitate general-purpose robust bird detection, agnostic to any specific application,
was key to how we designed the challenge specification.
The task of `detecting' birds in audio can be operationalised in multiple ways:
for example, a system that emits a trigger signal in continuous time representing the onset of each bird call,
a system that identifies regions of pixels in a spectrogram representation (time-frequency `boxes'),
or a system that estimates the number of calling individuals in a given time region \citep{Benetos:2017chapter}.
For any given application, the choice of approach will depend on the requirements for downstream processing.
We selected an option which we consider gave wide relevance,
while also being a task that could be solved by diverse methods,
from simple energy detection, through to template matching or machine learning.
This was that audio should be divided into ten-second clips,
and the task specification would be to label each clip with a binary label indicating the presence or absence of birds.

This approach quantises time such that any positive detection should be time-localisable within $\pm 10$ seconds,
which is sufficient for most purposes.
It also restricts such that there is no indication of the absolute number of bird calls detected within a positively-labelled clip;
however this is hard to ground-truth accurately.
Also via statistical ecology methods relative abundances may still be inferred from the distribution of positive detections \citep{Marques:2012}.
A concrete advantage of this approach was that it was much quicker to gather manual data annotations than would be the case for more complex labelling.

\subsection{Datasets}

We gathered and annotated datasets from multiple sources.
The purpose of this was twofold:
firstly to provide better evaluation of the generality of algorithms,
and secondly to provide challenge participants with development data
(e.g.\ to perform trial runs, or to train machine learning algorithms)
in addition to testing data.

We used audio data from remote monitoring projects and also from crowdsourced audio recordings.
These two dataset types differ from each other in many ways, for example:
remote monitoring audio was passively gathered, while crowdsourced audio recordings were actively captured;
the ratio of positive and negative items was different;
remote monitoring used fixed and known recording equipment, while crowdsourcing used uncontrolled equipment.
These differences were deliberately introduced for their use in ensuring that the challenge would be a strong test of generalisation.

\begin{description}
%%%%%%%%%%%%%%%%%%%%%%%%
\item[Chernobyl dataset:]

Our primary remote-monitoring dataset was collected in the Chernobyl Exclusion Zone (CEZ) for a project
to investigate the long-term effects of the Chernobyl accident on local ecology \citep{Wood:2016,Gashchak:2017}.
The project had captured over 10,000 hours of audio since June 2015, across various CEZ environments, using Wildlife Acoustics SM2 units.
For the present work we selected six of recording locations representing different environments (Table \ref{tbl:chernlocs}),
and from those selected a deterministic subsample: continuous 5-minute audio segments at hourly intervals, across multiple days.
Annotators manually labelled all time intervals in which birds were heard (using Raven Pro software),
and then we split recordings and metadata automatically into ten-second segments.
The number of files per location is uneven because of limited annotator time, giving us 6,620 items in total.
No weather filtering or other rejection of difficult regions was applied.

\begin{table}
  \centering
  \begin{tabular}{l | l l}
Codename	&	Habitat         	&	Radiation \\
\hline
Buryakovka	&	(TBC)             	&	Low \\
S2        	&	Deciduous forest	&	Medium \\
S11       	&	Meadow area     	&	Medium \\
S37       	&	Pine forest     	&	High \\
S60       	&	Shrub area      	&	Low \\
S93       	&	Mixed forest    	&	High \\
  \end{tabular}
  \caption{\label{tbl:chernlocs} Recording locations in Chernobyl dataset.}
\end{table}

%%%%%%%%%%%%%%%%%%%%%%%%
\item[Warblr dataset:]

Our first crowdsourced dataset came from a UK-wide project \textit{Warblr}.
Warblr is a software app available for Android and Apple smartphones,
which offers automatic bird species classification
(using the method of \cite{Stowell:2014b})
for members of the public via the submission of ten-second audio recordings.
We extracted a dataset of 10,000 audio files gathered in 2015--2016.
The audio files were thus actively collected, recorded on diverse mobile phone devices,
and likely to contain various human noise such as speech and handling noise.
No assumptions can be made that the data were a representative sample of geographic locations, weather conditions, or bird species.
Metadata for the files indicated that they covered all the UK seasons, many times of day (with a bias towards weekends and mornings)
and geographically spread all around the UK, with a bias toward population centres.

All recordings were selected that fell within the time window of available data, limited to a maximum of 10,000.
No selection or filtering of the data were performed beyond the self-selection inherent in crowdsourcing.

Although the data included automatic estimates of which bird species were present,
these were not precise enough to be converted to ground-truth data for the detection challenge.
We thus performed manual annotation, with each item being labelled as positive or negative according to the challenge specification.
Most items were single-annotated, although we were able to obtain double-annotation for a small number of items,
which allowed us to estimate inter-rater reliability.
Annotation was performed by experienced listeners using headphone listening and a simple web interface.

%%%%%%%%%%%%%%%%%%%%%%%%
\item[freefield1010 dataset:]

Our second crowdsourced dataset was an existing public dataset called \textit{freefield1010} \cite{Stowell:2014f}.
This consists of 7690 audio clips selected from the \textit{Freesound} online audio archive.
To create this dataset the audio clips had been selected such that they were labelled with the `field-recording' tag in the database,
and trimmed to ten seconds duration.
The data were of different origin than Warblr: they covered a global geographic range, and the recording devices used were almost never documented,
but likely to include hand-held audio recorders as used by pro-amateur sound recordists, as well as some mobile phones and some higher-end recording devices.
The Freesound database is crowdsourced and thus largely uncontrolled.

These data did not come with labels suitable for our challenge;
instead, each item came with a set of freely-chosen tags to indicate the content generally.
We investigated the `birdsong' tag, one of the most commonly used (2.6\% of items),
but found this insufficiently accurate.
We therefore had these audio annotated through the same process as the Warblr data.

%%%%%%%%%%%%%%%%%%%%%%%%
\item[PolandNFC dataset:]
The last dataset contains recordings from one author's (HP) project of monitoring autumn nocturnal bird migration. The recordings were collected every night, from September to November 2016 on the Baltic Sea coast, near Darlowo, Poland. We used Song Meter SM2 units with weather-resistant, directional Night Flight Calls microphones from Wildlife Acoustics Inc., mounted on 3--5 m poles. The amount of collected data (>3200 h of recordings) exceeded what human expert can annotate manually in reasonable time. Therefore we subjectively chose and manually annotated the subset consisting of 22 half-hour recordings from 15 nights with different weather conditions and background noise including wind, rain, sea noise, insect calls, human voice and deer calls. No other selection criterion or weather filtering was applied. Manual annotation was performed by visual inspection of a spectrogram and listening to the audio files. Only the passerine migrant calls were annotated (voices in 5--10 kHz range), so it may happen that some low pitched bird species (e.g.  resident owl calls) obtained no-bird label. However, such calls were extremely rare in described dataset, so this did not bias results significantly.

The selected recordings were then split into 1 s clips. The chosen clip length, different from other datasets, was chosen due to nocturnal bird calls typical duration (10--300 ms).

More details about the dataset structure (and analysis of the effect of audio clip duration) may be found in \cite{Pamula:2017}.

\end{description}
All sound files used in the public challenge were normalised in amplitude and saved as 16-bit single-channel PCM at 44.1 kHz sampling rate,%
\footnote{Normalisation via the \texttt{sox} tool using \texttt{gain -n -2}.}
and are available under open licences (see \textit{Data Accessibility} statement).

%\subsubsection{Post-hoc validation of testing set}

Our data annotation process was designed after early community discussions about how the challenge should be conducted.
We resolved that the annotations should reflect plausible annotation conditions as encountered in applications.
In particular, they should be well-annotated, yet any mislabellings discovered in the groundtruth data as the challenge progressed should not be eliminated,
since training data in practice do contain some errors and are not subject to the same scrutiny as in a data challenge.
A good detection algorithm must be able to cope with a small level of imprecision in the annotation data.

However, it was possible at the end of the challenge to perform further analysis and inspect the degree of machine errors and human errors.
To make good use of annotator time we used mismatch between automatically inferred decisions and manual annotations to search for mislabelled items in the dataset. For this we used the mean decision from the strongest three submissions to the challenge. All items in the testing set with a negative groundtruth label but a mean decision greater than 0.2, and all items with a positive groundtruth label but a mean decision less than 0.3, were examined and relabeled if needed. One might expect the threshold for re-validation to be 0.5: the asymmetry is because systems generally exhibited a bias towards low confidence, as will be seen later (Section \ref{sec:challoutcomes}). This re-validation process refined the testing set, but also allowed us to calculate a value for the inter-rater agreement for manual annotation, which we will express as an AUC for comparison against the results of automatic detection. Note that the re-validation process requires the time of expert listeners, and so it was not feasible to perform mass crowdsourcing on the whole collection.

\subsection{Baseline classifiers}

To establish baseline performance against which to compare new methods, we used two existing machine-learning based classification algorithms.

The first (code-named \textit{smacpy}) was the same baseline classifier as used in a 2013 challenge on ``detection and classification of acoustic scenes and events'' (``DCASE'') \citep{Stowell:2015}.
This baseline classifier represents a well-studied method used in many audio classification tasks: audio is converted to a representation called mel frequency cepstral coefficients (MFCCs), and the distributions of MFCCs are then modelled using Gaussian mixture models (GMMs).
Such an approach is simple, efficient and adaptable to many sound recognition tasks.
It has been superseded for accuracy in general-purpose sound recognition by more advanced methods \citep{Stowell:2015}.
We selected it to provide a common low-complexity baseline, and also because its simplicity meant it might successfully avoid \textit{overfitting} to the training data,
i.e.\ avoid becoming overspecialised, given that the training and test data would have different characteristics.

The second baseline (code-named \textit{skfl}) was a recent and more powerful classifier introduced for bird species recognition \citep{Stowell:2014b}.
This was the strongest audio-only bird species classifier in a 2014 international evaluation campaign.
Relative to smacpy it innovated in both the feature representation and the classification algorithm.
The feature representation was an automatically-learnt data transformation: two layers of ``unsupervised feature learning'' applied to Mel spectrogram input.
For classification the method used a random forest, an ensemble learning method based on decision trees that has emerged as powerful and robust for many tasks in machine learning \citep{Breiman:2001}.
Both of these components are known to work well with difficult classification scenarios, such as multi-modal classes, unbalanced datasets, and outliers.
We thus selected this second baseline as a representative of modern and flexible machine learning, designed for bird sounds.
In principle it could be more vulnerable to overfitting than the first baseline.
However, because of the inherent difficulty of the task, we expected skfl to perform more strongly than smacpy, and to provide a high performing baseline.

Python code for each of the baseline classifiers has previously been published \citep{Stowell:2014b,Stowell:2015}.
Initial results using the baseline classifiers were published online as a guide to the challenge participants.%
\footnote{\url{http://machine-listening.eecs.qmul.ac.uk/2016/10/bird-audio-detection-baseline-generalisation/}}

%%%%%%%%%%%%%%%%%%%%%%%%%%%%%%%%%
\subsection{The public challenge}

Conduct of the public challenge followed the design of previous successful contests on related topics \citep{Stowell:2015, Goeau:2016}. Having already determined the need for the challenge from remote monitoring literature and practitioners \citep{Stowell:2016c}, we announced intentions and led community discussion on the task design via a dedicated mailing list.

We collated our audio datasets into three portions:
development data to be publicly shared (7,690 items of freefield1010 plus 8,000 items from Warblr),
testing data whose true labels were to be kept private (10,000 items from Chernobyl plus 2,000 items from Warblr),
and a separate set not used for the challenge itself but for further study of algorithm generalisation (PolandNFC).
Providing two distinct development data sets allowed participants to test generalisation from one to the other, as part of their own algorithm development process,
while keeping some datasets fully private allowed us to evaluate generalisation without concern that algorithms might have been configured to the specifics of a given dataset.

The public development datasets were distributed in September 2016, both audio and groundtruth annotations. Teams could then begin to develop methods and train their systems. In December 2016 we released the testing data (audio only), with a one month deadline for the submission of inferred detection labels. The short time horizon of one month was intended to minimise the opportunity for overly adapting the methods to the characteristics of the testing data. During this period, teams could make multiple submissions, but limited to a maximum of one per day. "Preview" results, calculated from 15\% of the testing data, were provided in an interactive online plot, in order to give approximate feedback on performance (Figure \ref{fig:timelinescores}).

Participants were allowed to run their software on their own machines and then to submit merely the outputs (as opposed to the software code), which our online system would then score without revealing the groundtruth labels for the testing data.
Given that this open approach has potential vulnerabilities---such as recruiting manual labellers rather than developing automatic methods---we required the highest-scoring teams to send in their code which we inspected and re-ran on our own systems, to ensure a fair outcome.

Aside from intrinsic motivation, incentives for participants were cash prizes: one for the strongest scoring system , and one judges' award decided according to the use of interesting or novel methodology. This was done to stimulate conceptual development in the field, as opposed to the mere application of off-the-shelf deep learning. Participants were further required to submit technical notes describing their method, and later were invited to submit peer-reviewed conference papers to a special session at the European Signal Processing Conference (EUSIPCO) 2017. Of these, 8 challenge-related papers were accepted and presented.

The challenge organisation was thus designed to achieve the following: public benchmarking of methods against a common task and data, specifically tailored to fully-automatic configuration-free bird detection in unseen conditions; public documentation of the methods used to achieve leading results; and greater attention from machine learning researchers on data analysis tasks in environmental sound monitoring.

%\begin{figure}[h!]
%\centering
%\includegraphics[width=0.3\textwidth]{images/badch_data_partitioning}
%\caption{\label{fig:partitioningdiag}Diagram to illustrate the partitioning of data into development and testing data}
%\end{figure}

%%%%%%%%%%%%%%%%%%%%%%%
\subsection{Evaluation}

Our goal was to evaluate algorithms for their ability to perform general-purpose bird detection, within the selected format of binary decisions for ten-second audio clips.
A strong algorithm is one that can reliably separate the two classes ``bird(s) present'' and ``no bird present''.
However, since our evaluation was general and not targeted at a specific application,
we wished to generalise over the possible tradeoffs of precision versus recall (the relative cost of false-positive detections versus false-negative detections).
This strongly motivated our design such that participants should return probabilistic or graded outputs---a real-valued prediction for each audio clip rather than simply a 1~or~0---and our evaluation would use the well-studied \textit{area under the ROC curve} (AUC) as the primary quality metric.
The AUC measure has numerous qualities that make it well-suited to evaluation such classification tasks:
it generalises over all the possible thresholds that one might apply to real-valued detector outputs;
unlike raw accuracy, it is not affected by ``unbalanced'' datasets having an uneven mixture of positive and negative items;
chance performance for AUC is always 50\% irrespective of dataset;
and it has a probabilistic interpretation, as the probability that a given algorithm will rank a randomly-selected positive instance more highly than a randomly-selected negative instance \citep{Fawcett:2006}.

The ranking interpretation just mentioned highlights another aspect of the AUC statistic:
it treats detector outputs essentially as ranked values and is thus invariant to any monotonic mapping of the outputs,
in particular to whether the outputs are well-calibrated probabilities or not.
Well-calibrated, in this context, implies that when a detector outputs "0.75" for an item, this matches the empirical probability that in three out of four such cases the item is indeed a positive instance  \citep{Niculescu:2005}.

Thus AUC does not evaluate calibration.
But the need for calibration depends on the application: if a detector is being used to select a subset of strong detections, or to rank items for further manual inspection, there may be no need for calibration.
However if the detections are to be used in some probabilistic model e.g.\ for modelling a population distribution, it is desirable for a detector to output well-calibrated probabilities.
If a detector performs well in the sense evaluated by AUC, then its outputs can be mapped to probabilities by a post-processing step \citep{Niculescu:2005}.
Hence, we used AUC as our primary measure of quality, and separately we analysed the calibration of the submitted algorithms using the method of \textit{calibration plots},
which are histogram plots comparing outputs against empirical probabilities \citep{Niculescu:2005}.

%%%%%%%%%%%%%%%%%%%%%%%%%%%%%%%%%
\subsection{Further analysis via PolandNFC dataset}

After the challenge concluded we took the highest-scoring algorithm and applied it to the PolandNFC dataset,
an unseen and difficult dataset containing night flight calls, often brief and distant.
We used this in two ways:
(a) trained on a held-out portion (72.7 \%) of the PolandNFC data and tested on the remaining 27.3\%;
(b) trained using the main challenge development data and again tested on the 27.3\% of the PolandNFC data.

This allowed us to evaluate further the generalisation capability learnt by the network. For variant (a), the training dataset consisted of sixteen 30 min recordings collected over 11 nights (split into 28,784 1 s clips), and the testing dataset had six recordings from 4 nights (split into 10,793 1 s clips).
Training and testing recording dates were disjoint sets.
The testing set was held the same across variant (a) and (b) to ensure comparability of results.
Important to note is the specific structure of the PolandNFC dataset---it contains mostly negatively annotated examples (only 3.2 \%  for testing and 1.6 \% for training set were positive).

%%%%%%%%%%%%%%%%%%%%%%%%%%%%%%%%%%%%%%%%%%%
\section{Results}

%\subsection{Validation and baseline tests}

In re-validating the testing set we examined those items with the strongest mismatch between manual and automatic detection, to determine which was in error: 500 presumed negative and 1243 presumed positive items. This showed inter-rater disagreement in 16.6\% of such cases---predominantly, the most ambiguous cases with barely-audible bird sounds with amplitude close to the noise threshold. Note that this percentage is not representative of disagreement across the whole dataset, but only on the `controversial' cases. We also observed that a strong mismatch according to the automatic detectors did not necessarily imply human mislabelling: some perceptually obvious data items could be consistently misjudged by algorithms. We will discuss algorithm errors further below. Through re-validation the inter-rater reliability, measured via the AUC, was measured as 96.7\%. This value provides an approximate upper limit for machine performance since it reflects the extent of ambiguity in the data according to human listeners' perception.

\begin{figure}[t]
\centering
\includegraphics[width=0.6\textwidth]{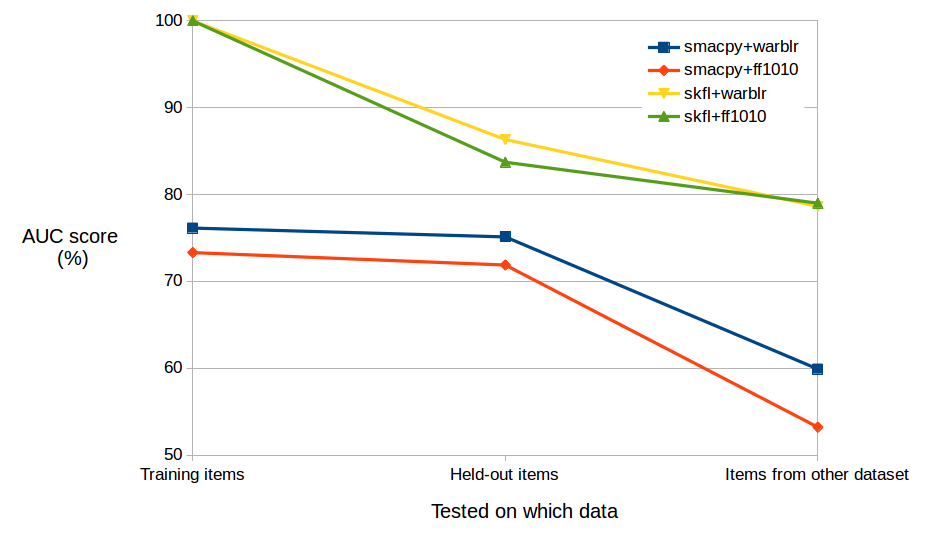}
\caption{\label{fig:baselinetests}Tests using baseline classifiers on the two development datasets. The middle column shows testing results under matched conditions, as commonly reported in machine learning; the rightmost column shows testing results under mismatched conditions, i.e. training with one dataset and testing with the other.}
\end{figure}

The two baseline classifiers gave relatively good performance on the development data, the strongest at over 85\% AUC in matched conditions, but generalised poorly. The simpler GMM-based baseline classifier showed consistently lower results than the more advanced classifier, as expected. It also showed strong resistance to overfitting in the sense that its performance on its training set was a very good predictor of its performance on a matched-conditions testing set. However, this was not sufficient to allow it to generalise to mismatched conditions, in which its performance degraded dramatically (Figure \ref{fig:baselinetests}). The more advanced baseline classifier also degraded when tested in mismatched conditions, though to a lesser extent, attaining 79\% AUC.

\subsection{Challenge outcomes}
\label{sec:challoutcomes}

\begin{figure}[ht]
\centering
\includegraphics[width=0.7\textwidth,page=4]{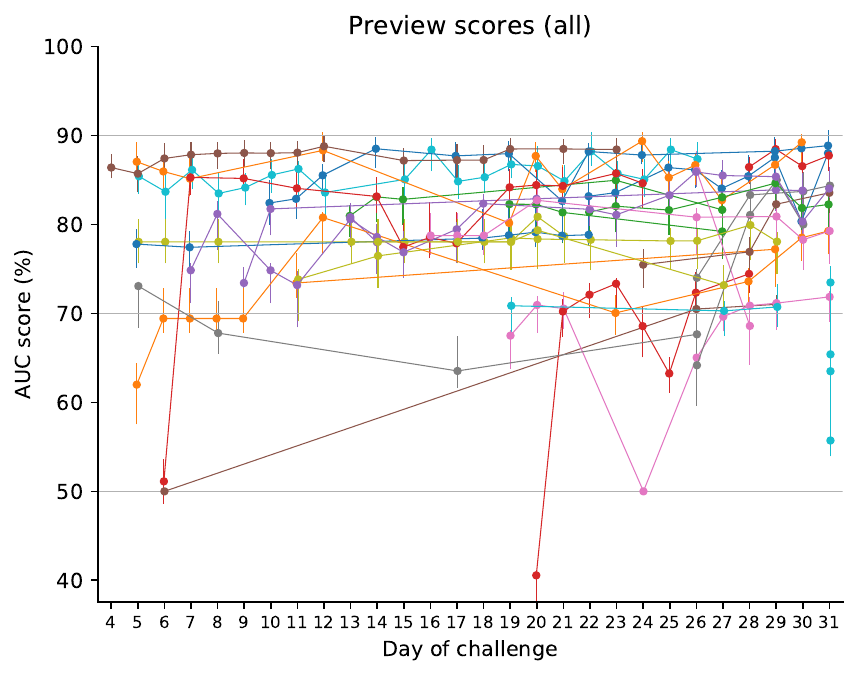}
\caption{\label{fig:timelinescores}Timeline of preview scores attained by every submission for each team. Error bars are estimated by bootstrap sampling.}
\end{figure}

\begin{figure*}[p]
% * <szarotkaaa@gmail.com> 2018-02-17T08:52:28.078Z:
% 
% Shouldn't we add a legend with teams names? Maybe couple best submissions, e.g. four? Or you think that we don't need to as it's not a point of this figure. 
% 
% ^.
\centering
\begin{sideways}
\centering
\includegraphics[width=0.95\textheight]{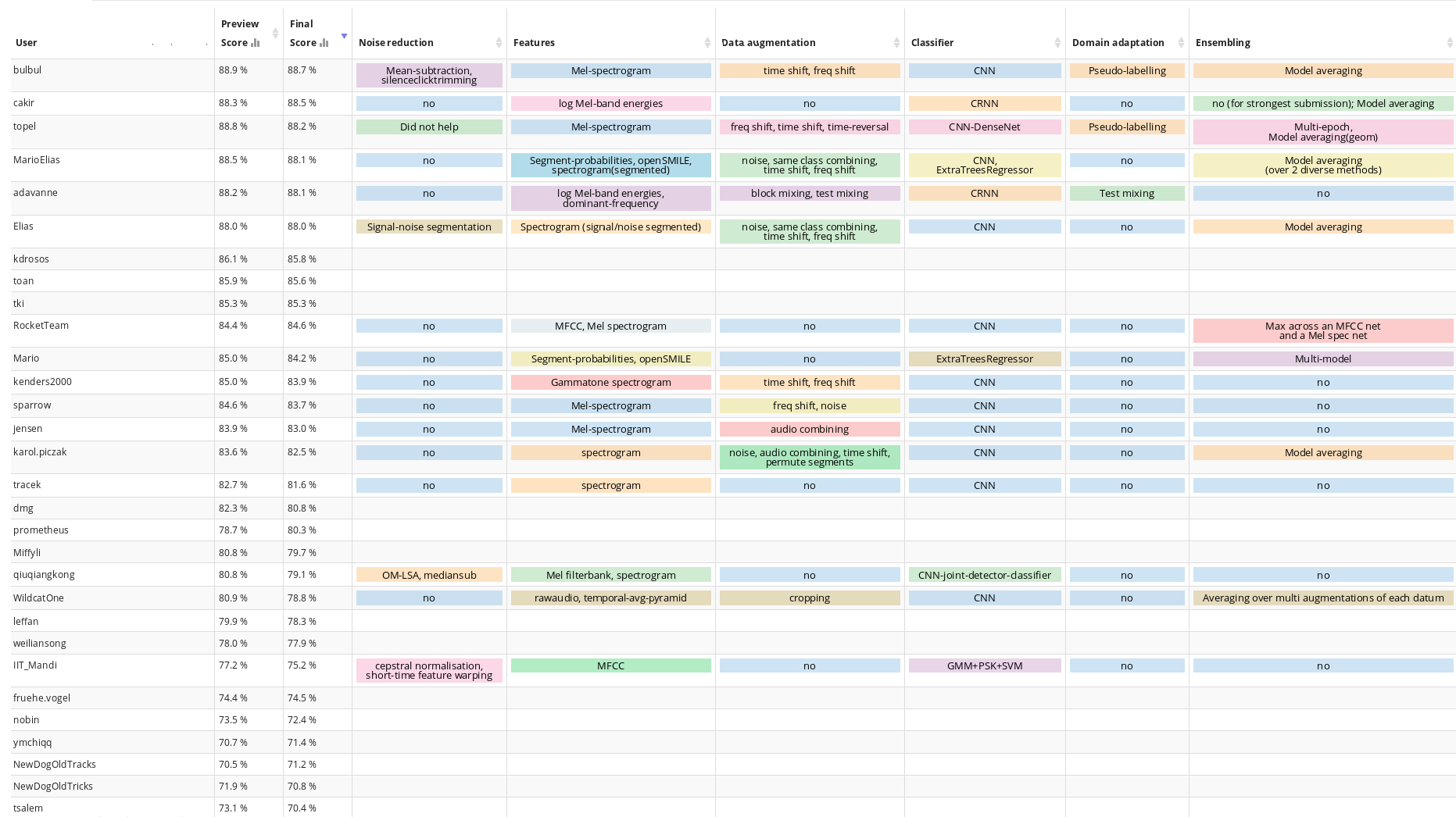}
\end{sideways}
\caption{\label{fig:resultstbl}Overview of the methods used by the submitted systems, and their scores.}
\end{figure*}

Thirty different teams submitted results to the challenge, from various countries and research disciplines, with many submitting multiple times during the one-month challenge period (Figure \ref{fig:timelinescores}).
Around half of the teams also submitted system descriptions, of which the majority were based on deep learning methods, often convolutional neural networks (CNNs) (Figure \ref{fig:resultstbl}).
To preprocess the audio for use in deep learning, most teams used a spectrogram representation, often a "mel spectrogram", which is a spectrogram with its frequency axis warped to an approximation of human nonlinear frequency-band sensitivity.
Many teams also used data augmentation, meaning that they artificially increased the amount of training data by copying and modifying data items in small ways, such as adding noise or shifting the audio in time.
These strategies are in line with other work using machine learning for bird sound \citep{Goeau:2016,Salamon:2017}.

%%%%%%% TODO COULD DROP THIS FIGURE

\begin{figure}[th]
\centering
\includegraphics[width=0.7\textwidth,page=3]{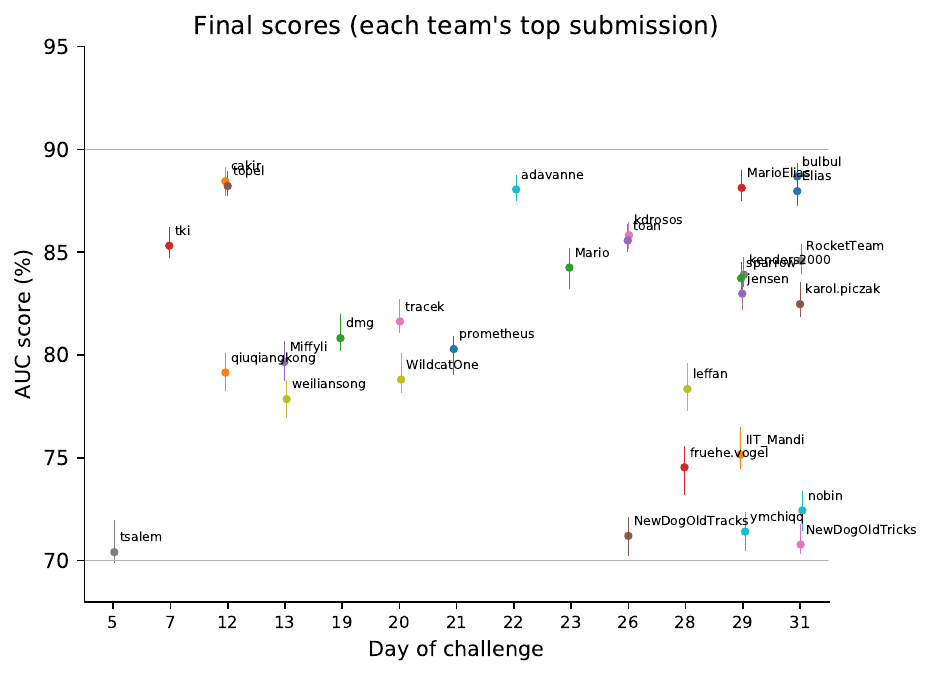}
\caption{\label{fig:finalscoresall}Final scores attained by the highest-performing submission for each team. Error bars are estimated by bootstrap sampling.}
\end{figure}

\begin{figure}[th]
\centering
\includegraphics[width=0.7\textwidth,page=6]{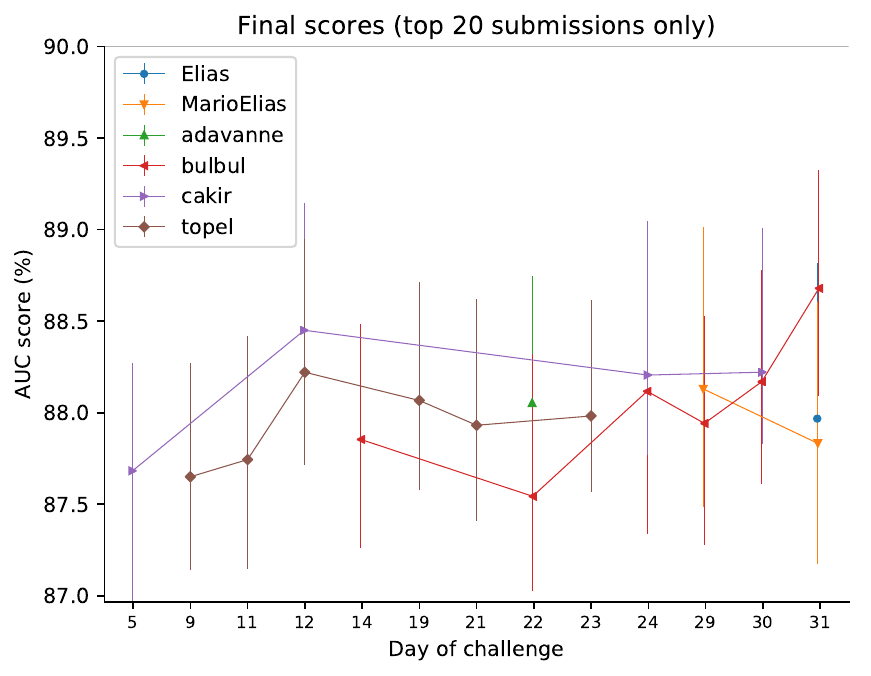}
\caption{\label{fig:finalscorestop}Final scores attained by the 20 highest-scoring submissions. Error bars are estimated by bootstrap sampling.}
\end{figure}

Most teams were able to achieve over 80\% AUC, but none over 90\%:
the strongest score was 88.7\% AUC, attained by team `bulbul' (Thomas Grill) on the final day of challenge submission (Figure \ref{fig:finalscoresall}).
The team has given further details of their approach in a short conference publication \citep{Grill:2017} and with open-source code available online.%
\footnote{\url{https://jobim.ofai.at/gitlab/gr/bird_audio_detection_challenge_2017/tree/master}}
Inspecting the highest-scoring entries, and estimating their variation through bootstrap sampling, we found that four teams' results were within the confidence interval of the highest score (Figure \ref{fig:finalscorestop}).

\begin{figure}[h!]
\centering
\includegraphics[width=0.4\textwidth,page=11]{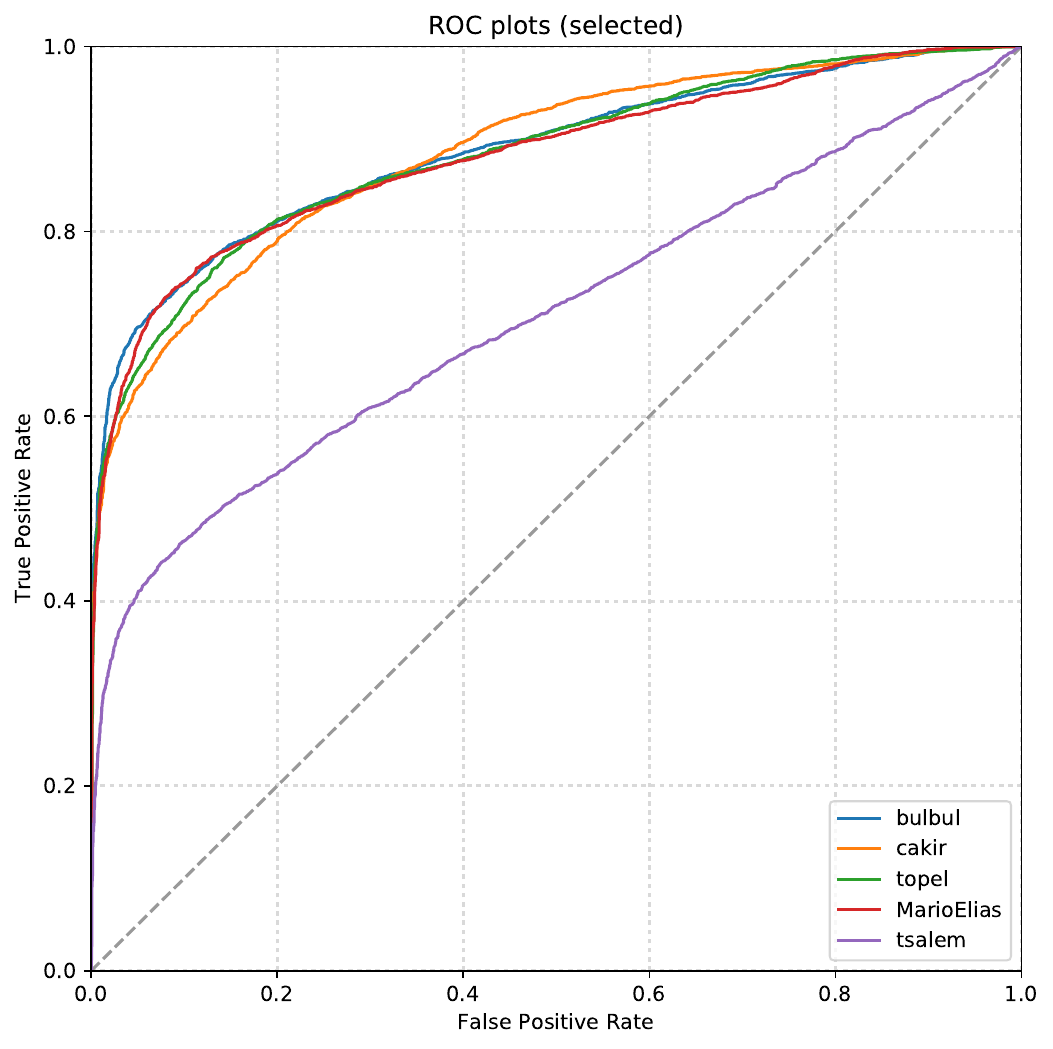}
\caption{\label{fig:rocplots}ROC plots for the systems attaining the four highest and one lowest AUC score.}
\end{figure}

AUC scores are summary statistics of ROC plots.
Inspecting the detail of the ROC plots for these four highest-scoring systems and for the lowest-scoring system (Figure \ref{fig:rocplots}),
we find some tendency for the curves to be biased towards an asymmetry (close to the left-hand edge of the plot but not the upper edge).
This implies a spread of `difficulty' for the test items: there were some positive items that were easy to detect without incurring extra false-positives as a side-effect,
while many remained difficult to detect.
The most balanced of the ROC plots inspected was that of `cakir', implying that this system had a more balanced distribution of its discriminative power across the easy and difficult cases.

\begin{figure*}[h!]
\centering
\includegraphics[width=0.48\textwidth,page=12]{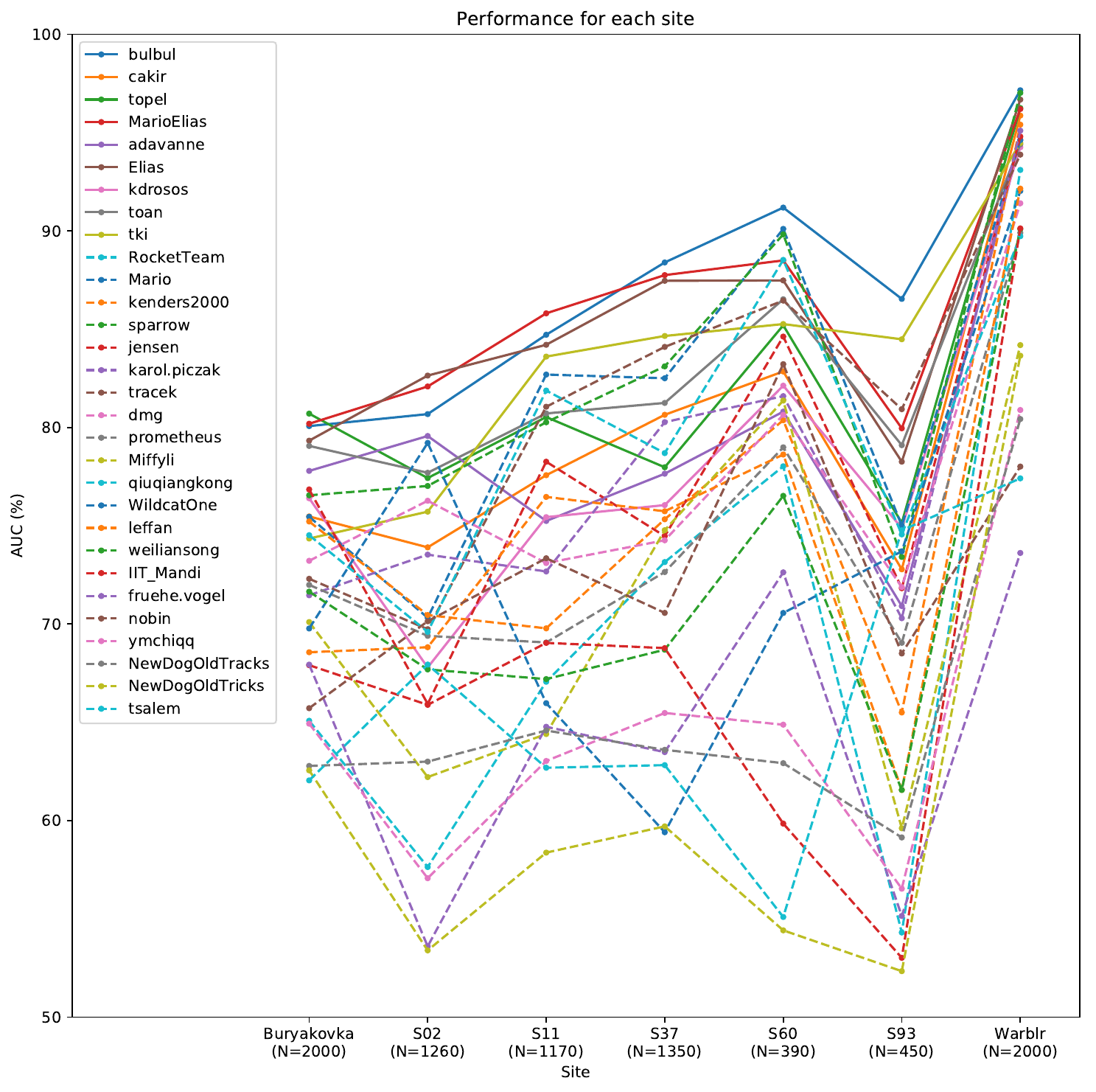}
\includegraphics[width=0.48\textwidth,page=13]{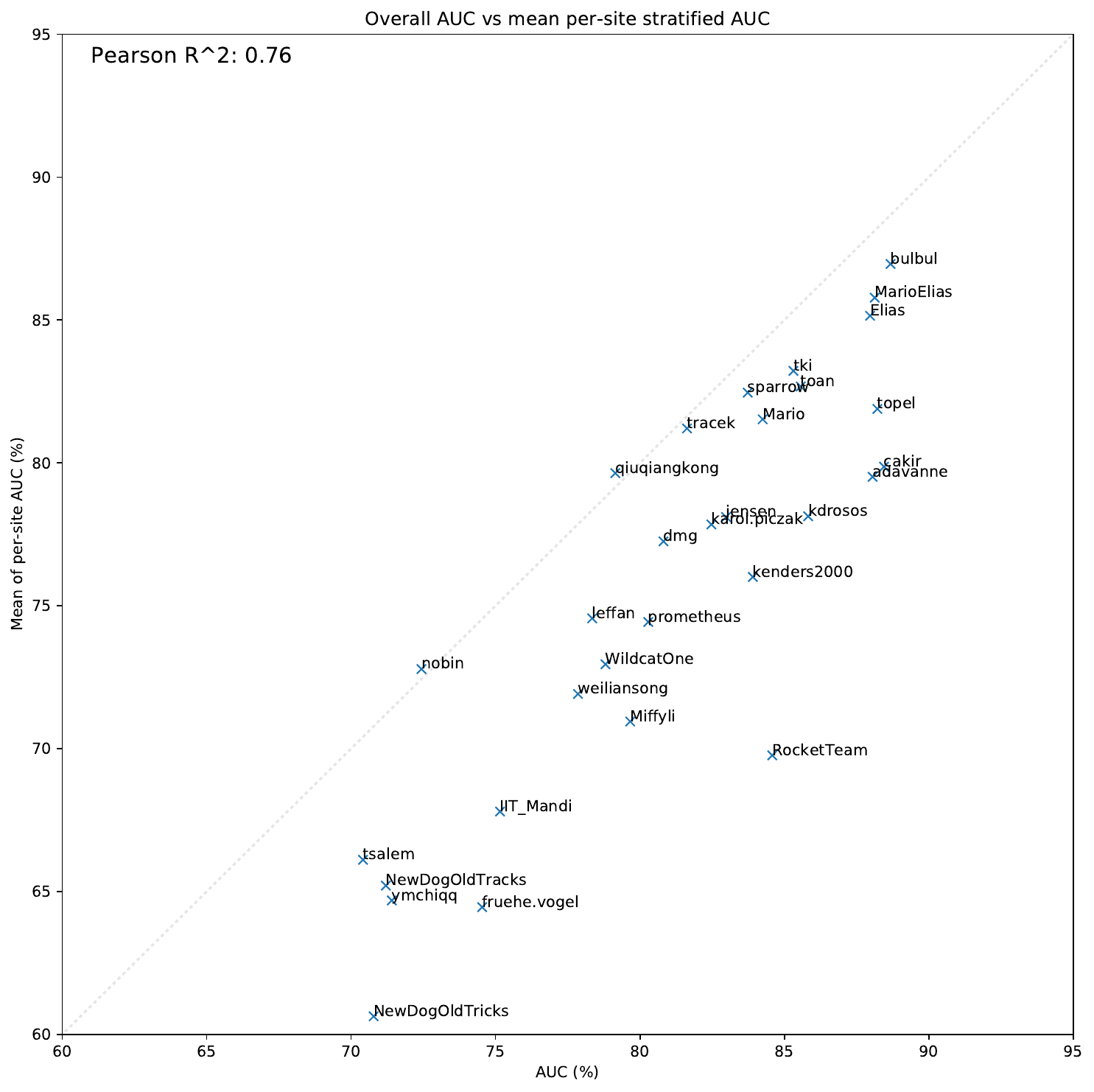}
\caption{\label{fig:aucpersite}Performance (AUC) of submitted algorithms analysed on a per-site basis. Warblr data (from around the UK) is treated as one site, while the other sites are recorders in the CEZ. Left-hand plot shows the per-site results; right-hand plot shows how the AUC scores compare when calculated over the whole pooled dataset, versus as a mean of the stratified per-site AUCs.}
\end{figure*}

Since the testing data consisted of items from multiple `sites'---i.e.\ known sites in the Chernobyl Exclusion Zone, plus the Warblr (UK) data considered as a separate single site---we were able to calculate the AUC scores on a per-site basis (Figure \ref{fig:aucpersite}, left).
These showed a strong site-dependency of algorithm performance.
Whereas Warblr data could be detected with an AUC of over 95\%,
Chernobyl sites showed varying difficulty for the detectors overall, some as low as 80\% even for the leading algorithms.
However, the overall AUC was very highly predictive of the average of the per-site AUCs (Pearson $R^2 = 0.76$; Figure \ref{fig:aucpersite}, right).
Note that the two AUC calculations are not independent and so some correlation is expected.
The observed correlation validates that the overall AUC is usable as a summary of the per-site performance.
The `bulbul' system remained the strongest performer even under the per-site analysis.
The rank ordering of systems was not highly preserved: for example the second-placed `cakir' system would have been ranked tenth if using the average per-site AUC.

\begin{figure}[h!]
\centering
\includegraphics[width=0.7\textwidth,page=7]{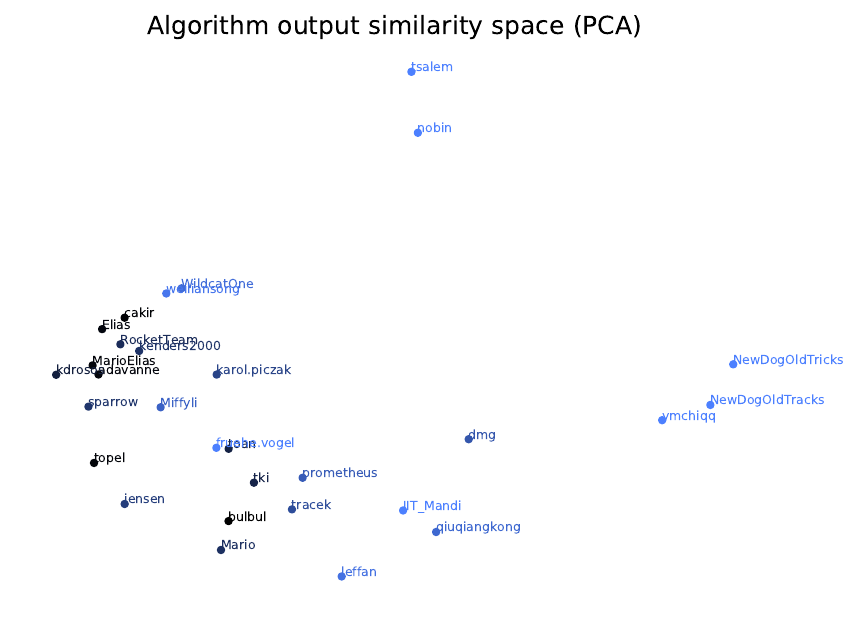}
\caption{\label{fig:pcaspace}Similarity space comparing the top-scoring submission by each team (PCA projection of the submitted predictions after rank-transformation). Submissions are close together if their predictions were similar, irrespective of their accuracy. Submissions obtaining higher AUC scores are darker in colour.}
\end{figure}

Applying PCA to the system outputs, we recovered an abstract `similarity space' of systems, which showed that some of the lowest-scoring systems formed two outlier clumps in terms of their predictions, while the stronger systems formed something of a continuum (Figure \ref{fig:pcaspace}).
The very strongest scoring systems did not cluster tightly together, indicating that there remains some diversity in the strategies implicit in these high-performing detectors.

\begin{figure}[h!]
\centering
\includegraphics[width=0.48\textwidth,page=9]{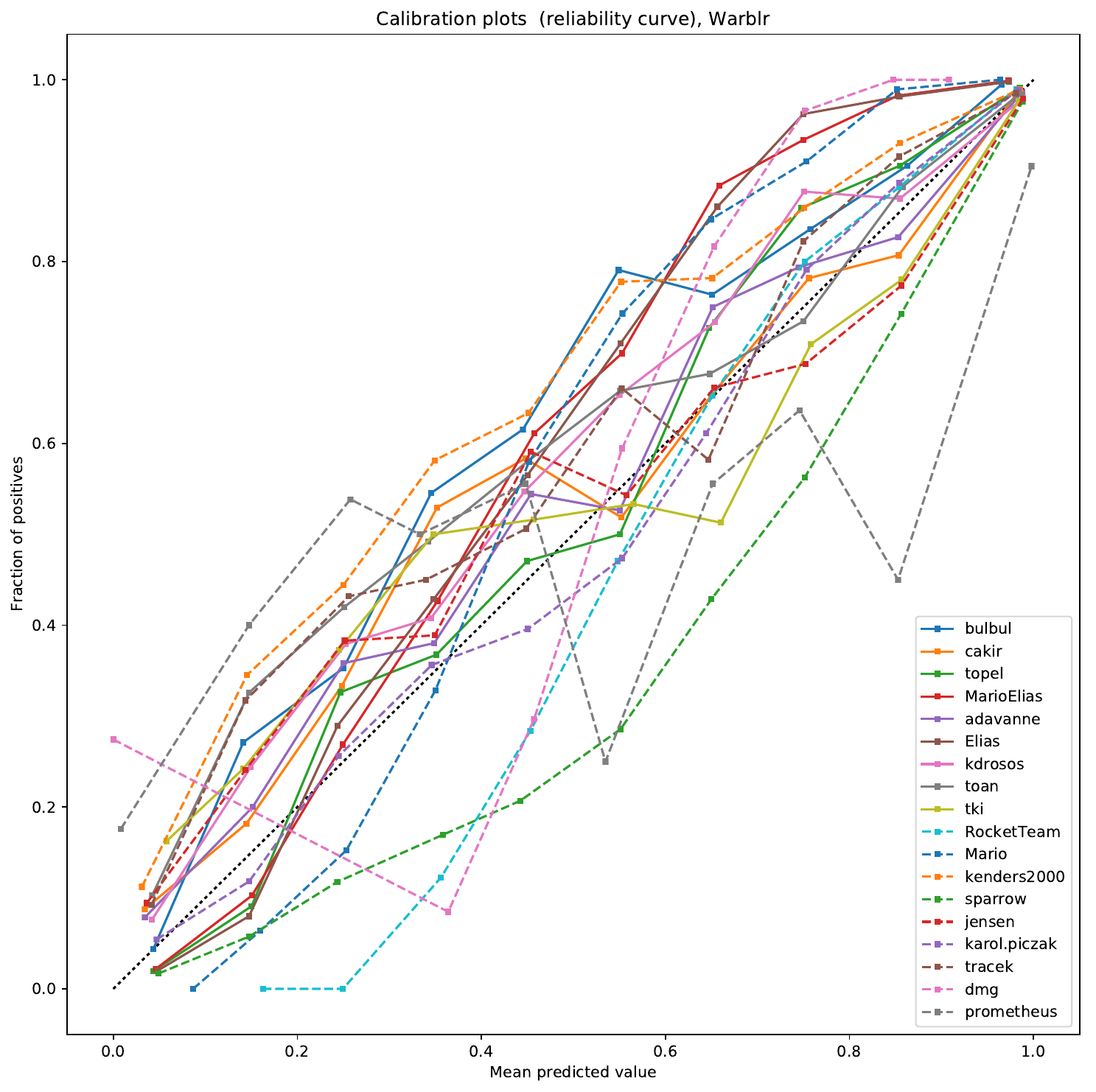}%
\includegraphics[width=0.48\textwidth,page=10]{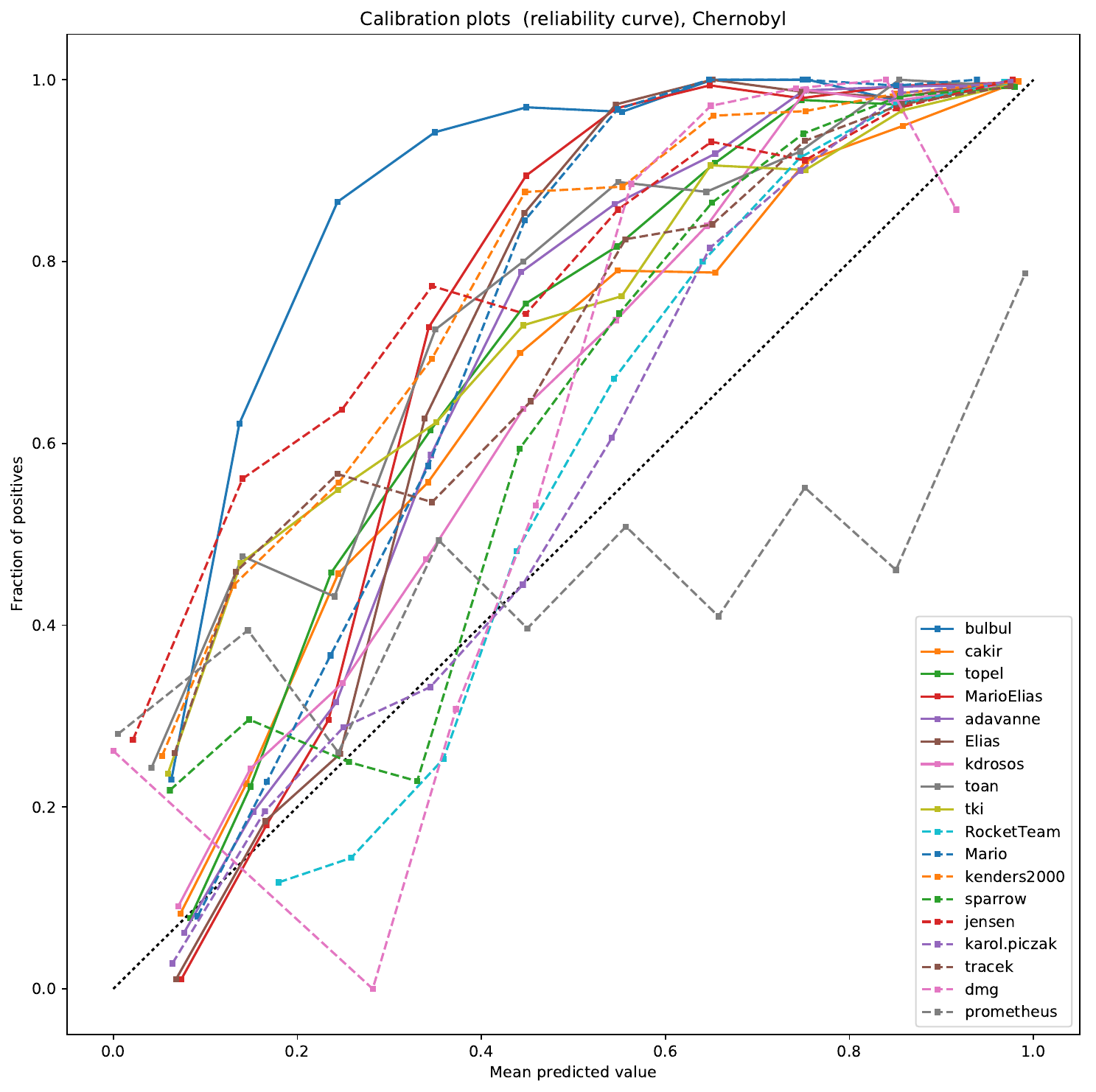}
\caption{\label{fig:calibrationplot}Calibration plots for the strongest submission by each team, separately for the Warblr test data (first plot) and Chernobyl test data (second plot). For legibility we have limited this to the submissions attaining at least 80\% AUC. A submission whose outputs are well-calibrated probabilities should yield a line close to the identity diagonal.}
\end{figure}

We measured calibration curves separately for the Warblr and Chernobyl testing data (Figure \ref{fig:calibrationplot}).
Calibration was generally better for Warblr, as one might expect given the availability of Warblr training data.
Notably, the highest-scoring submission `bulbul' had by far the worst calibration on the Chernobyl data:
around 80\% of cases it assigned a prediction value of 0.25 were indeed positive (versus around 30\% on the Warblr data).
The second-highest-scoring submission `cakir' exhibited quite different behaviour, remaining relatively well calibrated even when assessing the unseen Chernobyl data.

%%%%%%%%%%%%%%%%%%%%%%%%%%%%%%%%%%%%%%%%%%%%%%%%%%%%%%%%%%%%%%%%%%%%%%%%%%%%%%%%%%%%%%%%%%%%
\subsection{Error analysis}

\begin{table}
  \centering
  \begin{tabular}{l | r r}
Category	&	False positives	&	False negatives \\
\hline
\hline
Clear                                  	&	1	&	68 \\
Dontknow                               	&	7	&	40 \\
\hline
\hline
Faint (e.g. v distant)                 	&	0	&	179 \\
Short call                             	&	0	&	69 \\
Noise-masking (inc wind, river)        	&	0	&	67 \\
Insect                                 	&	26	&	52 \\
Human (speech, laughter, tv, imitation)	&	31	&	13 \\
Rain (inc drops)                       	&	26	&	5 \\
Unusual bird sound                     	&	0	&	29 \\
Misc distractor                        	&	0	&	11 \\
Misc mammal                            	&	2	&	0 \\
  \end{tabular}
  \caption{\label{tbl:errorcats} Inferred reasons for mistakes made by the strongest-performing systems, annotated for the 500 items for which the systems showed the strongest deviation from groundtruth. Note that the count data sum to more than 500, since multiple reasons could potentially be attributed to each item. ``Clear'' means the item was perceptually clear and should have been correctly labelled; ``Dontknow'' means that no obvious reason for a mistake is evident, even if the item is not particularly clear; all other rows are categories of presumed reasons for machine errors.}
\end{table}

We inspected the 500 data items for which the predictions of the strongest systems exhibited mismatch with the re-validated ground truth,
to characterise typical errors made by even the strongest machine learning systems in bird audio detection (Table \ref{tbl:errorcats}).
Such inspection is heuristic, relying on perceptual judgment to estimate the causes of errors;
however repeated tendencies give us indications about the performance of the current state of the art.

For false negatives, by far the most common observation was that positive items contained very faint bird sound (e.g.\ distant), often needing multiple listens to be sure it was present.
These faint sounds had low SNR and were often also quite reverberated.
Low SNR was also a factor in the third most common presumed cause, noise masking.
This category included general broadband `pink' noise sources including wind and rivers.
There were other more specific categories of sound that appeared to act as masker or distractor causing systems to overlook the bird sound: insect noise was common in the CEZ data, while human sounds such as speech, whistling, or TVs were present in the Warblr data.
The second most common presumed cause of false negatives was however the presence of extremely short calls: often a single ``chink'' sound, which might perhaps be overlooked or confused with rain-drop sounds.
Some sounds were presumed to be missed because they were unusual for the dataset (e.g.\ goose honking), although this was not seen as a major factor.

False positives occurred at a much lower rate in the top 500 most mismatched items. They appeared to be caused in roughly equal proportion by insects, human sounds, and rain sounds (individual drops or diffuse rainfall).

%%%%%%%%%%%%%%%%%%%%%%%%%%%%%%%%%%%%%%%%%%%%%%%%%%%%%%%%%%%%%%%%%%%%%%%%%%%%%%%%%%%%%%%%%%%%
\subsection{Further analysis via PolandNFC dataset results}

\begin{figure}[t]
\centering
\includegraphics[width=0.48\textwidth,page=1,clip,trim=0mm 0mm 0mm 7mm]{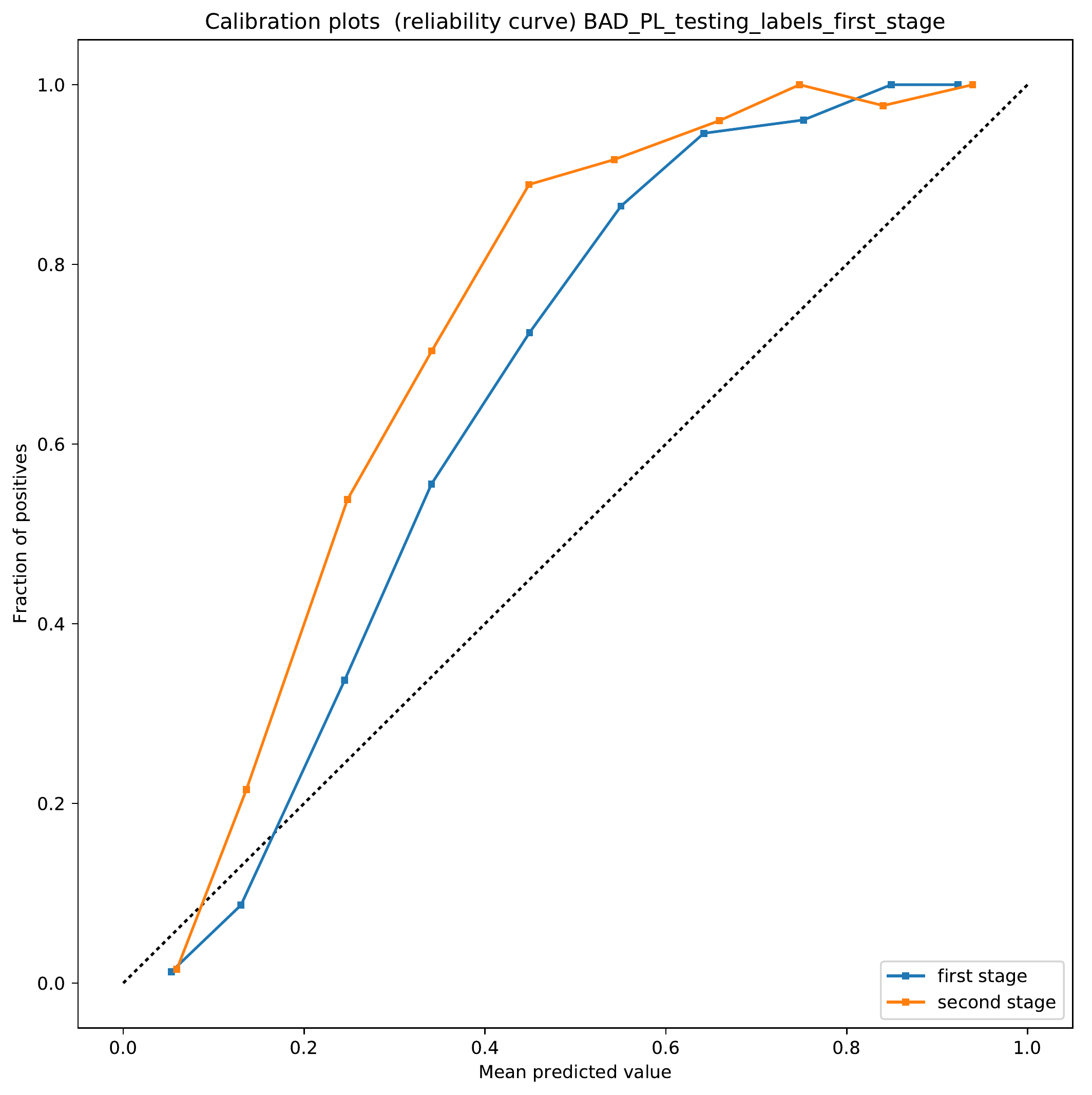}
\includegraphics[width=0.48\textwidth,page=2,clip,trim=0mm 0mm 0mm 7mm]{images/plot_badchall_hannadata}
\caption{\label{fig:plcalib}Calibration plots obtained when applying the highest-scoring challenge system to remote monitoring audio data from Poland (cf.\ Figure \ref{fig:calibrationplot}). First plot: Challenge (mismatched) training; second plot: PolandNFC (matched) training.}
\end{figure}

We then applied the highest-scoring method (`bulbul') to the separate and unseen acoustic monitoring dataset PolandNFC.
The bulbul algorithm performed its inference in two stages:
the first stage applied the pre-trained neural network to make initial predictions;
the second stage allowed the neural network to adapt to the observed data conditions,
by feeding back the most confident predictions as new training data \citep{Grill:2017}.
We evaluated the outputs from each stage (Table \ref{tbl:plauc}).
AUC results were of high quality and were most strongly affected by the choice of training data,
the matched-conditions training yielding much more accurate predictions.
The second stage adaptation offered some improvement in the case where the training set came from mismatched conditions.
However in matched-conditions training the second stage actually incurred a slight reduction in performance.
Calibration plots for this test show that the detector was also better calibrated when trained in matched conditions, and that the second stage re-training did not have a strong effect on calibration (Figure \ref{fig:plcalib}).

\begin{table}[t]
  \centering
  \begin{tabular}{ l | r r}
              & \multicolumn{2}{c}{AUC (\%)} \\
Training data & First stage  &  Second stage \\
\hline
Challenge (mismatched) & 83.9  (+/-  0.4)& 87.76 (+/-  0.07) \\ 
PolandNFC    (matched) & 94.95 (+/-  0.01) & 93.78 (+/-  0.01)\\
  \end{tabular}
  \caption{\label{tbl:plauc} Performance obtained when applying the highest-scoring challenge system to remote monitoring audio data from Poland, average of results from two runs of the system.}
\end{table}
% * <szarotkaaa@gmail.com> 2018-02-17T09:14:56.417Z:
% 
% I've replaced the results by average of two runs, only for first result the range is big enough to have one decimal place 
% 
% ^.

%%%%%%%%%%%%%%%%%%%%%%%%%%%%%%%%%%%%%%%%%%%
\clearpage
\section{Discussion}

%\subsubsection{* Overall discussion}

Two broad observations emerge from this study:

\begin{enumerate}
\item
Machine learning methods, primarily deep learning, are able to achieve very high recognition rates on remote monitoring acoustic data, despite weather noise, low SNRs, wide variation in bird call types, and even with mismatched training data. % (Fig 4, 6, 7).
The AUC results presented here are a dramatic advance in the state of the art, and machine learning methods are of practical use in remote monitoring projects.
\item
However, there remains a significant gap between performance in matched conditions and in mismatched conditions. % (Fig 7, 9, 10, Table 3).
True generalisation remains difficult, and further work is needed.
Projects are thus recommended to obtain some amount of matched-conditions training data where possible,
and to treat automatic detection results with some caution especially with regard to the calibration of the outputs if they are to be treated as probabilities or if a fixed detection threshold is used.
(Post-processing such as Platt scaling can ameliorate calibration issues \citep{Niculescu:2005}.)
If a ranked-results approach is used (e.g. keeping the strongest $N$ detections), which circumvents questions of calibration, then performance can remain strong even in mismatched conditions (Fig 7; Table \ref{tbl:plauc}).
\end{enumerate}

In practical applications there are differing tradeoffs of precision versus recall, of false negatives versus false positives.
The AUC statistics summarise over these but tell us that whatever position on the ROC curve is chosen, the improvement in the state of the art which we report can give a much better tradeoff (Fig 6), which corresponds for example to a much lower amount of manual postprocessing time in filtering out false-positive results \citep{Pamula:2017}.

The strongest machine learning methods in this study were convolutional and/or recurrent neural nets (CNNs, RNNs, or CRNNs), as has been observed in other domains \citep{Lecun:2015}.
In order to ensure methods could work in conditions different from those in the training data,
various participants explored self-adaptation, in which a trained network is fine-tuned upon exposure to the new conditions (without needing any additional ground-truth information)
\citep{Grill:2017,Cakir:2017,Adavanne:2017}.
Participants reported mixed results of this, some observing no benefit.
We found little benefit of self-adaptation for matched conditions;
however, in cases where matched-conditions training data is not available,
we found that it can reduce the adverse effect of the mismatch (Table \ref{tbl:plauc}).

A further practical question is the feasibility of implementation on low-power devices for long-term deployment in the field.
Deep learning experiments often require hardware acceleration, primarily for the training phase.
After training, deep learning algorithms can be deployed onto smaller embedded units \citep{MacAodha:2018}.
However, the self-adaptation methods considered here are essentially additional rounds of training, albeit conducted with unlabelled data,
and thus would incur quite some cost for use in the field.
A pragmatic version of this would be to perform training or `pre-training' using mismatched data,
then collecting a small amount of matched data from the target field conditions to perform self-adaptation,
before fixing the algorithm parameters for use on-device.

%\subsubsection{* Error analysis}

Our error analysis showed that the most common error for the present generation of algorithms appears to be false-negatives due to the failure to detect faint/distant bird sound.
The next most salient concern is robustness to masking noise (wind, river, insects, speech).
There were many entries which were very hard to decide, as a listener: faint, short, and masked sounds together constitute a large portion of items.
This perceptual difficulty, reflected in inter-rater disagreements, reminds us that some cases may be inherently ambiguous and thus may always be difficult for machine recognition.
However, there is still a fair proportion (more than 1 in 10 of the top 500 items inspected) for which the sound recording was judged to be perceptually clear,
meaning that the reason for those false negatives is due to a detector failing to model bird sound correctly, providing scope for algorithm improvements.

A further specific cause of detector errors stems from the ambiguity between very short `chink' bird calls and sounds such as individual rain drops which have similar effects in a spectrogram.
A related issue was observed in bats, with the very short calls of species in the \textit{Myotis} genus being the most difficult to disambiguate according to \citet{Walters:2012}.
If the observable attributes of multiple sources overlap entirely then it is not possible to distinguish them even in principle.
However, at least in our case human listeners can tell the difference, whether from context or from fine detail of signals. 
We thus expect that future work on higher-resolution input features---such as waveform data rather than spectrogram data---will be able to improve on this issue.

\citet{Hutto:2009} previously performed an analysis of human sound detection of birds. Their comparison was between humans and ``autonomous recording units'':
however, note that in the latter case the detection was performed manually by inspecting spectrograms and listening to recordings, contrasted against a human listener in the field.
Their results are thus not directly comparable to ours; however they too found that distant bird sounds were the predominant cause of missed detections for remote sensing units.
\citet{Furnas:2015} likewise studied in-field versus audio-based detection using manual annotation, with similar results. They noted that detection probability could vary according to situational factors such as elevation and tree canopy cover.
\citet{Digby:2013} compared evaluated automated detection in audio against in-field manual detection, for a single species (the little spotted kiwi, \textit{Apteryx owenii}), finding detection rates of around 40\% with a relatively simple detection algorithm; despite this, they concluded that the high efficiency of automatic methods led to a large reduction in person-hours and thus was recommended. They found that wind noise exerted a larger influence on automatic detection than on manual detection.

Overall, our study design via a data challenge has been successful in moving forward the state of the art in acoustic remote monitoring.
The design as a binary classification task, evaluated by AUC, is recommended as a way to generalise over some diversity in requirements among remote monitoring projects,
with the calibration analysis as an important addition to AUC evaluation.
The use of multiple test sets sourced from different projects is a robust approach for general-purpose evaluation of algorithms,
and we further recommend the use of per-site stratified AUC to account for per-site differences.%
\footnote{The second edition of the Bird Audio Challenge, launched at time of writing, incorporates these recommendations, using per-site stratified AUC as well as adding further test sets to the challenge. \url{http://dcase.community/challenge2018/task-bird-audio-detection}}
This complements the task-specific evaluation that a well-resourced individual project should undertake (cf. \citet{Knight:2017}).
Going beyond the ``yes/no'' binary classification task, in some cases is desirable to identify individual bird calls:
the binary classification paradigm can in fact enable this, through a procedure of ``weakly supervised learning'' \citep{Morfi:2017deductive,Morfi:2018multitask,Kong:2017}.
In future evaluations we recommend the exploration of such approaches, combining broad-scale detection with the elucidation of finer detail.

%%%%%%%%%%%%%%%%%%%%%%%%%%%%%%%%%%%%%%%%%%%
\clearpage
\section*{Acknowledgments}

We would like to thank
Paul Kendrick for preparation and annotation of Chernobyl data,
Luciana Bar\c{c}ada for annotation of Chernobyl data,
and Julien Ricard for programming and administering the challenge submission website.
We also thank the many challenge participants for their enthusiastic effort.

\section*{Author contributions}

DS, MW, YS and HG designed the data challenge study and its analysis;
DS, MW and HP provided datasets;
HG led the creation of the challenge submission website;
DS and HP performed tests of machine learning systems;
DS, HG and HP analysed the results;
DS led the writing of the manuscript, with some sections by HP.
All authors contributed critically to the drafts and gave final approval for publication.

%%%%%%%%%%%%%%%%%%%%%%%%%%%%%%%%%%%%%%%%%%%
\section*{Data accessibility}
\begin{itemize}
\item Development datasets: both audio and annotations are available under CC-BY-4.0 licences.
\\ warblrb10kaudio: \url{https://archive.org/details/warblrb10k_public}
\\ ff1010bird audio: \url{https://archive.org/details/ff1010bird}
\\ Annotations: \url{https://doi.org/10.6084/m9.figshare.3851466.v1}
\item Testing data (Chernobyl/warblrb10k): audio public, but annotations held back for future challenges.
\\ Audio: \url{https://archive.org/details/birdaudiodetectionchallenge_test}
\item PolandNFC data: held back, in preparation for future challenge.
\item Source code for baseline classifiers: \\ GMM `smacpy' classifier (MIT licence): \url{https://github.com/danstowell/smacpy}.
\\ skfl feature-learning is available as supplementary information in \citet{Stowell:2014b} (MIT licence).
\item Source code for the online submission website (MIT licence): \url{https://github.com/julj/submission}
\item Source code and papers for various of the systems submitted by challenge teams are available via \url{http://c4dm.eecs.qmul.ac.uk/events/badchallenge_results/} (various licences).
\end{itemize}

%%%%%%%%%%%%%%%%%%%%%%%%%%
%\bibliographystyle{plainnat}
%\bibliographystyle{IEEEtran}
%\bibliography{../../refs}
\bibliography{references}

\end{document}